\documentstyle[aps,epsfig]{revtex}
\begin{document}
\sloppy
\title{Polyelectrolyte Networks: Elasticity, Swelling, and
  the Violation of the Flory - Rehner Hypothesis 
\footnote{Dedicated to Bruce Eichinger. The authors consider it a great 
honor to be invited to contribute
     to the special issue for Bruce Eichinger. }}
\author{ T.A. Vilgis
   and J. Wilder}
\address{Max-Planck-Institut f\"ur Polymerforschung,
Postfach 3148, 55021 Mainz, Germany}
\date{\today}
\maketitle

\begin{abstract}
This paper discusses the elastic behavior of polyelectrolyte networks. The
deformation behavior of single polyelectrolyte chains is discussed. It is
shown that a strong coupling between interactions and chain elasticity
exists. The theory of the complete crosslinked networks shows that the Flory
- Rehner - Hypothesis (FRH) does not hold. The modulus contains
contributions from the classical rubber elasticity and from the electrostatic
interactions. The equilibrium degree of swelling is estimated by the assumption
of a $c^{*}$-network.
\end{abstract}
\vspace{1cm}


\section{Introduction}

The elastic behavior of polyelectrolyte networks is not understood.  The main
reason is the lack of knowledge in the interplay between the elastic part
$F_{\rm el}$ of the free energy and the part form the interactions $F_{\rm
  int}$.  The simplest approximation to overcome this problem is the
application of the Flory - Rehner - Hypothesis (FRH) which suggests that both
parts of the free energy can be added \cite{fh}. Therefore the thermodynamic
behavior of the network is then determined by the minimum of the total free
energy $F = F_{\rm el}+F_{\rm int}$. This procedure must be wrong in general
as pointed out by Ball and Edwards \cite{ball}. The first reason is purely
energetic. If the partition function is formulated by using the Edwards
technique (see \cite{edvil} for a review), it must be expected that a cross
term between both contributions exists. On the other hand it was shown
recently that for realistic networks with fractal structural units, 
"left overs" from the network formation processes yield immediately a
violation of the FR - hypothesis \cite{sommer}. These results have been shown
by analytical considerations  and by numerical simulations.

Clearly the violation of the FRH  was here shown to have purely
structural reasons and it can be argued that for perfect networks the FRH
should hold, as it was believed to be confirmed by many experimental work 
\cite{exp,exp1} on neutral networks and conventional rubbers upon 
welling experiments. Indeed it is difficult to say anything on the validity of
the FRH in swelling of conventional rubbers. Real rubber samples cannot be
characterized as they should be and only indirect measurements such as
neutron scattering \cite{neutron} can give some hints, but the conclusions in
favor or against the FRH seem to be vague. For a clear statement on the
validity of the FRH, classical rubbers suffer from the fact that the
interactions in these systems are weak. The interactions are "only" excluded
volume forces and the corresponding potential is very short ranged. The short
ranged excluded volume potential means that Flory Huggins type,
or Random Phase Approximations (see for example \cite{degennes,doi}) work very
well. However, the excluded volume interactions are weak and thus 
their effect on
the elastic properties is indeed difficult to study with swelling experiments
only. 

Polyelectrolyte networks do not suffer from the lack of weak
interactions. Moreover the range of the interaction can be changed during
experiments. Very often the interaction between chain segments are described
by a Debye - H\"uckel potential of the form $V({\bf r}) \propto 1/r \exp(-
\kappa r)$, where $r$ is the spatial distance between two segments and
$\kappa$ the Debye screening parameter, which depends mainly 
on the total ionization $I$
\cite{degennes,tanford,schmitz}, i.e., $\kappa^{2} \propto I$. 
The latter means
that the total ionization can be changed easily by adding and removing salt
to / from the solution. The range and the effect of the interactions can be
changed easily. Large ionization $I$ corresponds a high concentration of salt
and thus to large values of $\kappa$. In this case the interaction potential
can be safely modeled by a short ranged form, in some sense with an excluded
volume type potential with a certain value determined by the Bjerrum length
and $\kappa$ \cite{borsali}. The more interesting case is therefore the regime
of small values of the Debye screening parameter corresponding to long range
potentials. This is the regime discussed below in more detail. 

We will show below systematically that the FRH is not applicable to the case
of polyelectrolyte networks. To do so we first derivate the Flory-Rehner
Hypothesis. Then to make the paper self contained we
 start with the computation of the deformation of a single polyelectrolyte
chain, though some progress has already been achieved \cite{rubi} by the
application of blob-pictures \cite{degennesblob}.  
Indeed these ideas have been also applied to networks \cite{rubi}. The problem
is here is that the such obtained results are based on single chain
approaches. Therefore these results are restricted to cases where either the
meshes are largely stretched, or the electrostatic interactions are
screened. These two cases are reflected on the ratio of the screening length
and the chain size, i.e., $1/\kappa$ much larger than the mean mesh size
radius the network or vice versa. In the latter the network behaves almost as
a classical swollen network, whereas the first case corresponds to the strong
polyelectrolyte regime \cite{rubi}. In the present paper we will also present
a theory beyond these two extreme regimes. Moreover we will leave in the
following the single mesh approximation and study a network composed by many
meshes. Then we cannot rely on the extreme cases $\kappa$ large or small..   

Our results will show that there is a strong interplay between the
elastic force and the interactions. The force extension relationship is in
contrast to neutral Gaussian chains not only dependent on the chain length and
temperature \cite{treloar} but also a strong effect on the single chain
modulus from the electrostatic interaction appears \cite{HaWiVi}. This
method will be generalized to polyelectrolyte networks \cite{wilder}. We
show below that the elastic modulus is a combination of both contributions,
i.e., the classical elastic part, determined by crosslinks only \cite{deam} 
and a new part coming from the interactions. Thus the modulus is determined
strongly by the strength and the range of the interaction. 
Note that this is not
the case for neutral networks. All computations for excluded volume networks 
so far show, that the modulus is determined by the number of crosslinks
only \cite{deam,PanRab}. In the third main part of the paper we use the single
chain results to study the swelling behavior of polyelectrolyte
networks. Before we proceed we present a  mathematical formulation of the
origin of the FRH to show what has to be done to go beyond the classical
theories.

\section{Derivation and Limits of the Flory-Rehner-Hypothesis}
We consider a polyelectrolyte network in solvent containing $N_{\rm i}$
indistinguishable  counter-ions. The partition function for such a network
consisting of one very long single chain
with a fixed crosslink-configuration ${\bf S}$ in terms of a path integral reads:
\begin{eqnarray}
Z({\bf S})=\frac{1}{N_{\rm i}!}\int {\cal D}{\bf r}(s)\int \frac{{\mbox d}{\bf
  R}_{1}}{V_{0}}\dots\int \frac{{\mbox d}{\bf R}_{N_{\rm i}}}{V_{0}}\exp(-\beta
H)\prod_{(i,j)}\delta[{\bf r}(s_{i})-{\bf r}(s_{j})],\label{frh1}  
\end{eqnarray} 
here $(i,j)$ denotes that the $i$-th and the $j$-th monomer form a
crosslink, ${\bf r}(s)$ represents the chain conformation in three dimensions as a function of the
contour variable $s$ and ${\bf R}_{l}$ is the position of the $l$-th
counterion. $V_{0}$ denotes the integration-volume.  As we assume that the network is weakly charged the underlying
Hamiltonian $H$ is given by an Edwards Hamiltonian for flexible
chains. Considering the counterions explicitly the Hamiltonian $H$ is:
\begin{eqnarray}
\beta H&=&{3 \over 2l^{2} } \int^{N_{0}}_{0} \mbox{d} s \
\left({\mbox{d} {\bf r} \over \mbox{d}
    s}\right)^2+\frac{\beta}{2}\int^{N_{0}}_{0}\mbox{d} s \int^{N_{0}}_{0}\mbox{d}
s'\frac{z^{2}}{4\pi\epsilon_{0}\epsilon_{\rm r}  \vert {\bf r} (s) - {\bf r} (s')
  \vert}\nonumber \\
&+&\beta\sum_{i=1}^{N_{\rm i}}\int^{N_{0}}_{0}\mbox{d} s\frac{zz_{\rm
    i}}{4\pi\epsilon_{0}\epsilon_{\rm r}  \vert {\bf r} (s) - {\bf R}_{i}
  \vert}+\frac{\beta}{2}\sum_{i,l=1}^{N_{\rm
    i}}\frac{z_{i}z_{l}}{4\pi\epsilon_{0}\epsilon_{\rm r}  \vert {\bf R}_{i} - {\bf R}_{l} \vert}\label{frh2}   
\end{eqnarray} 
where $l$ is the Kuhn length, $N_{0}$ the total number of monomers of the
crosslinked chain, $\beta$ is
$(k_{\rm B}T)^{-1}$, where $k_{\rm B}$ is the Boltzmann constant and $T$
denotes the absolute temperature. $\epsilon_{0}$ is the dielectric
constant and $\epsilon_{\rm r}$ the relative dielectric constant. $z$ is the
monomer charge in units of $e$ and $z_{i}$ is charge of the $i$-th counterion,
which is assumed to be independent of $i$.
${\bf R}_{i}$ denotes the position of the $i$-th counterion.
\newline
In order to use collective coordinates, we introduce the densities
$\rho_{\rm p}$ of the polymers and $\rho_{\rm i}$ for the counterions. The constrained
partition function for fixed densities $\rho_{\rm p}$ and $\rho_{\rm i}$ reads
(see Appendix A): 
\begin{eqnarray}
Z(\rho_{\rm p},\rho_{\rm i},{\bf S})&=&\exp\left[-\frac{1}{2}\sum_{\bf
    k}\left(\frac{1}{{\tilde S}_{0}({\bf k})}+V_{\rm pp}({\bf
      k})\right)\rho_{\rm p}({\bf k})\rho_{\rm p}(-{\bf k})\right]\nonumber \\
&\times&\exp\left[-\frac{1}{2}\sum_{\bf
    k}\left(\frac{1}{C_{0}({\bf k})}+V_{\rm ii}({\bf
      k})\right)\rho_{\rm i}({\bf k})\rho_{\rm i}(-{\bf k})-\sum_{\bf
    k}V_{\rm pi}({\bf k})\rho_{\rm p}({\bf k})\rho_{\rm i}(-{\bf
    k})\right]\label{frh10} 
\end{eqnarray}
with $V_{\rm pp}({\bf k})=bz^{2}/{\bf k}^{2}$. $V_{\rm pi}$ and $V_{\rm ii}$
are defined analogously. ${\tilde S}_{0}({\bf k})$ and $C_{0}({\bf k})$ are
defined as:
\begin{eqnarray}
{\tilde S}_{0}({\bf k})=\left<\int_{0}^{N_{0}} {\mbox d}s \int_{0}^{N_{0}}{\mbox d}s'\,{\rm
    e}^{i{\bf k}({\bf r}(s)-{\bf r}(s'))}\right>_{0}\label{frh7}
\end{eqnarray}
and
\begin{eqnarray}
C_{0}({\bf k})=\left<\sum_{j,l=1}^{N_{\rm i}}{\rm
    e}^{i{\bf k}({\bf R}_{j}-{\bf R}_{l})}\right>_{0}\label{frh8}
\end{eqnarray}
where $<\dots>_{0}$ means the expectation value with respect to $H_{0}$:
\begin{eqnarray}
\beta H_{0}={3 \over 2l^{2} } \int^{N_{0}}_{0} \mbox{d} s \
\left({\mbox{d} {\bf r} \over \mbox{d}
    s}\right)^2+\ln\left(\prod_{(i,j)}\delta[{\bf r}(s_{i})-{\bf
    r}(s_{j})\right)\label{frh9}
\end{eqnarray}
Eq. (\ref{frh10}) defines an effective Hamiltonian $H_{\rm eff}$:
\begin{eqnarray}
\beta H_{\rm eff}(\rho_{p},\rho_{i})&=&\frac{1}{2}\sum_{\bf
    k}\left(\frac{1}{{\tilde S}_{0}({\bf k})}+V_{\rm pp}({\bf
      k})\right)\rho_{\rm p}({\bf k})\rho_{\rm p}(-{\bf k})+\frac{1}{2}\sum_{\bf
    k}\left(\frac{1}{C_{0}({\bf k})}+V_{\rm ii}({\bf
      k})\right)\rho_{\rm i}({\bf k})\rho_{\rm i}(-{\bf k})\nonumber \\
&+&\sum_{\bf
    k}V_{\rm pi}({\bf k})\rho_{\rm p}({\bf k})\rho_{\rm i}(-{\bf
    k})\label{frh11}
\end{eqnarray}
From the effective Hamiltonian we separate the ${\bf k}=0$ term, i.e.:
\begin{eqnarray}
\beta H_{\rm eff}=\beta H^{0}_{\rm eff}(\rho_{p}(0), \rho_{i}(0))+\sum_{k>0}\beta
H'_{\rm eff}(\rho_{p}({\bf k}), \rho_{i}({\bf k}))\label{frh12}
\end{eqnarray}
with
\begin{eqnarray}
\beta H^{0}_{\rm eff}(\rho_{p}(0), \rho_{i}(0))&=&\frac{1}{2}\left(\frac{1}{{\tilde S}_{0}(0)}+V_{\rm pp}(0)\right)\rho_{\rm p}(0)\rho_{\rm p}(0)+\frac{1}{2}\left(\frac{1}{C_{0}(0)}+V_{\rm ii}(0)\right)\rho_{\rm i}(0)\rho_{\rm i}(0)\nonumber \\
&+&V_{\rm pi}(0)\rho_{\rm p}(0)\rho_{\rm i}(0)\label{frh13}
\end{eqnarray}
For pure Coulomb potentials 
 in infinite large systems the values for $V_{\rm pp}(0)$, $V_{\rm pi}(0)$ and
$V_{\rm ii}(0)$ appear to be
divergent. In realistic systems the chains (and meshsizes) are finite. The
lowest value for the wave vector is then of the order of  $|{\bf k^{2}}| \sim
1/(l^{2}N_{0})$ which acts as a
cut off for small wave vectors. Thus divergences are prevented. 

If fluctuations are neglected the only contribution to the partition function
stems from $H^{0}_{\rm eff}$. In this case the Free Energy $F_{0}$ reads:
\begin{eqnarray}
F_{0}&=&-\ln\left(Z_{0}(\rho_{\rm p}(0),\rho_{\rm
    i}(0))\right)\label{frh14} \\
&=&\frac{1}{2}\left(\frac{\rho_{\rm p}^{2}(0)}{{\tilde
      S}_{0}(0)}+\frac{\rho_{\rm
      i}^{2}(0)}{C_{0}(0)}\right)+\rho_{i}(0)\ln(\rho_{i}(0))+\frac{1}{2}V_{\rm pp}(0)\rho_{\rm p}^{2}(0)+\frac{1}{2}V_{\rm ii}(0)\rho_{\rm i}^{2}(0)+V_{\rm pi}(0)\rho_{\rm p}(0)\rho_{\rm i}(0)\nonumber
\end{eqnarray}
In deformed systems the first term of the right-hand-side of
Eq. (\ref{frh14}), which is written in brackets,
contains the elastic Free Energy. The second term represents the translational
entropy of the counterions and the last three terms are interaction parts. This equation indicates, that neglecting fluctuations the Free Energy is just
a sum of an elastic, an entropic and an interaction part 
$F=F_{\rm el}-TS_{\rm counter}+F_{\rm int}$. This reflects the   
Flory-Rehner Hypothesis. It shows up only on the limit ${\bf k} =  0$. Below
we will show that this is not all all sufficient. 
In general fluctuations cannot be neglected. In the present network
case fluctuations have a
strong  effect: they violate the Flory-Rehner Hypothesis and renormalize the
Gaussian elasticity. If the Free Energy
is calculated without neglecting the fluctuations one has to integrate over
all $\rho_{\rm p}({\bf k})$ and $\rho_{\rm i}({\bf k})$ by performing the path
integrals. Eq. (\ref{frh10}) shows that in this case there will be a strong
coupling between the elastic and the interaction part of the Free Energy. Even
in a Debye-H\"uckel approximation, i.e. $\rho_{\rm p}(k>0)=0$ this coupling
cannot be avoided. In the following sections we show, that in
polyelectrolyte networks there is a
coupling between the elastic and the interaction part and thus the Flory-Rehner
Hypothesis is no longer applicable. 

\section{Single Polyelectrolyte Chain}
To approach the problem of elasticity in polyelectrolyte networks on a systematic
basis we recall the results of the single chain deformation. This section will
introduce the mathematical method and - more important - it will be shown that
the force - extension relationship is fundamentally changed from what is known
in Gaussian and excluded volume chains.  The deformation of the end - to end
- distance of a Gaussian chain by an external force $f$ is simply described by 
\begin{equation}
\label{sc1}
f = k_{\rm B}T \frac{r}{l^{2}N} 
\end{equation}
where $T$ is the temperature, $l$ the Kuhn length, $N$ the chain length and
$r$ the distance of the chain ends. This result can be easily derived from the
Gaussian statistics of the ideal chain \cite{treloar}. Already when excluded
volume forces are taken into account eq.(\ref{sc1}) is changed in a fundamental
way. Excluded volume force change the statistical behavior of the chain, i.e.,
the size if the chain is given by $R \simeq  lN^{{3/5}}$. Thus the single
chain is swollen, and another stretching law is expected. This problem was
solved by Pincus \cite{degennes,Lpincus}. The result is briefly quoted. The
problem is no longer isotropic and the force has two components, i.e., parallel
to the stretching direction and perpendicular to the stretching direction. At low
deformation a Hookian deformation behavior is expected, whereas at larger
deformations  the self avoiding statistics comes into play. In the low
deformation regime the result is
\begin{eqnarray}
&\displaystyle\label{i1}
\vert \langle {\bf r}_{\parallel} \rangle \vert
\sim
\langle {\bf R}^2 \rangle f&
\end{eqnarray}
and in the limit $f \to \infty$
\begin{eqnarray}
&\displaystyle\label{i2}
\vert \langle{\bf r}_{\parallel} \rangle \vert
\sim
N f^{1/\nu-1}&
\end{eqnarray}
for the elongation in parallel direction  to the force ${\bf f}$.
For the perpendicular direction it is found ($f \gg 1$)
\begin{eqnarray}
&\displaystyle\label{i3}
\langle {\bf r}^2_{\perp} \rangle
\sim
N f^{1/\nu-2}&
\end{eqnarray}
A similar situation will occur in the case of polyelectrolyte chains. The
coupling between elasticity and interactions will become much stronger. The
interactions are very strong and pre-stretch the chain to a large extend. Thus
it is instructive  to study the elastic behavior of single polyelectrolytes.

\subsection{Model}

Let us first introduce the standard model which is employed here. Since we
restrict ourselves to flexible chains which are weakly
charged the Edwards model \cite{Ledwards} is the appropriate tool, like already given in
section II.
Here we consider a Debye-H\"uckel chain, thus neglecting fluctuations of the counterions:
\begin{eqnarray}
&\displaystyle\label{model1}   \beta H_{\rm E} [{\bf r};{\bf f}] = {3 \over 2l^{2} } \int^{N_{0}}_{0} \mbox{d} s \
\left({\mbox{d} {\bf r} \over \mbox{d} s}\right)^2 + \beta
 \int^{N_{0}}_{0} \mbox{d} s \  {\bf f} {\mbox{d} {\bf r} \over \mbox{d} s}
& \nonumber \\&\displaystyle
 + \
\frac{bz^{2}}{2} \int^{N_{0}}_{0} \mbox{d} s \int^{N_{0}}_{0} \mbox{d} s' \
{\exp\left\{-\kappa\left \vert {\bf r} (s) - {\bf r} (s') \right \vert \right \}
\over \left \vert {\bf r} (s) - {\bf r} (s') \right \vert}\,,
\end{eqnarray}
where  $b=e^{2}/4\pi
\epsilon_{0}\epsilon_{\rm r}k_{\rm B}T$ is the Bjerrum-length, $e$ is the
electrical charge of an electron, $\epsilon_{0}$ is the dielectric
constant and $\epsilon_{\rm r}$ the relative dielectric constant. ${\bf f}$ is the external force and $\kappa^{-1}$
denotes the Debye-H\"uckel screening length.
The correlation function can be calculated in terms of a path integral
\cite{Ledwards,Lfeynman,Lfreed} as follows:
\begin{eqnarray}
&\displaystyle\label{model2}
G\left({\bf r},N_{0};{\bf f}\right) =
\int^{{\bf r}(N_{0}) = {\bf r}}_{{\bf r}(0) = {\bf 0}}
{\cal D} {\bf r}(s) \
\exp\left\{- \beta H_{\rm E}[{\bf r};{\bf f}]\right\}&.
\end{eqnarray}
Its Fourier transform is defined by
\begin{eqnarray}
&\displaystyle\label{model3}
G({\bf k},N_{0};{\bf f}) = \int^{ }_{ } \mbox{d}^3 {\bf r}
 \
\exp\{-i{\bf k} {\bf r} \} G\left({\bf r},N_{0};{\bf f}\right)&.
\end{eqnarray}
The averages of the
force-size relationship $\langle{\bf R}^2_\parallel\rangle$ and
$\langle{\bf R}^2_\perp\rangle$,
where ${\bf R}_\parallel$ denotes the parallel component
with respect to ${\bf f}$ and ${\bf R}_\perp$ is the
corresponding perpendicular part, are then readily
calculated by the general formulae
\begin{eqnarray}
\langle{\bf R}^2_\parallel\rangle[{\bf f}]
=
- \left. { \partial^2/\partial k^2_{3} \ G({\bf k},N_{0};{\bf f})
\over G({\bf k},N_{0};{\bf f})}\right\vert_{{\bf k}={\bf 0}}\label{model4}&
\end{eqnarray}
and
\begin{eqnarray}
&\displaystyle\label{model5}
\langle{\bf R}^2_\perp\rangle[{\bf f}]
=
- \left. {\sum^{2}_{i=1} \partial^2/\partial k^2_i \ G({\bf k},N_{0};{\bf f})
\over G({\bf k},N_{0};{\bf f})}\right\vert_{{\bf k}={\bf 0}}&.
\end{eqnarray}
By analytic continuation of the Fourier space to the complex plane, the
correlation function $G({\bf k},N_{0};{\bf f})$ can also be written as the zero-force
correlation function $G({\bf k}-i\beta {\bf f},N_{0};{\bf f}={\bf 0})$.
Substitution of Eq.  \ref{model2} into Eq. \ref{model3} yields:
\begin{eqnarray}
 G({\bf k},N_{0};{\bf f})=\int^{ }_{ } \mbox{d}^3 {\bf r}\ 
\exp\{-i{\bf k} {\bf r} \}
\int^{{\bf r}(N_{0}) = {\bf r}}_{{\bf r}(0) = {\bf 0}}
{\cal D} {\bf r}(s)\ \exp \left \{-\beta H_{\rm E}[{\bf r};{\bf f}]\right \}
\label{model10}
\end{eqnarray}
 For constant force ${\bf f}$
Eq. (\ref{model10})  can be rewritten as:
\begin{eqnarray}
 G({\bf k},N_{0};{\bf f})&=&\int^{ }_{ } \mbox{d}^3 {\bf r}\ 
\exp\{-i({\bf k}-i\beta {\bf f}) {\bf r} \}
\int^{{\bf r}(N_{0}) = {\bf r}}_{{\bf r}(0) = {\bf 0}}
{\cal D} {\bf r}(s)\ \exp \left \{-\beta H_{\rm E}[{\bf r};{\bf f}={\bf 0}]\right \}
\nonumber \\
&=&  G({\bf k}-i\beta {\bf f},N_{0};{\bf f}={\bf 0})\label{model8}
\end{eqnarray}
Consequently, to get results for  $\langle{\bf R}^2_\parallel\rangle$ and
$\langle{\bf R}^2_\perp\rangle$, we only have to calculate $G({\bf k},N_{0})$ and continue the
first argument of $G$ to the complex plane.

\subsection{Calculation of the Correlation Function}

For technical  reasons it is convenient to introduce a field theory, which
enables to carry out the variational technique. Thus
the field theoretical Hamiltonian reads \cite{HaWiVi}:
\begin{eqnarray}
\beta H[\vec{\psi}]&=&\frac{1}{2}\int_{{\bf k}}\vec{\psi}(-{\bf k})\left[ \mu_{0}
  +\frac{l^{2}}{6}k^{2}\right]\vec{\psi}({\bf k})\label{ft3} \\&+&\frac{(2\pi)^{3}}{8}\int_{{\bf k_{1}},{\bf
    k_{2}},{\bf k_{3}},{\bf k_{4}}}\vec{\psi}({\bf k_{1}})\vec{\psi}({\bf k_{2}})U({\bf
  k_{1}}+{\bf k_{2}})\delta({\bf k_{1}}+{\bf k_{2}}+{\bf k_{3}}+{\bf
  k_{4}})\vec{\psi}({\bf k_{3}})\vec{\psi}({\bf k_{4}})\nonumber
\end{eqnarray}
where $\int_{{\bf k}}$ is an abbreviation for $\int d^{3}{\bf k}/(2\pi)^{3}$ and
 $U$ denotes the Debye-H\"uckel potential in units of $\beta^{-1}$.
In the Fourier space $\tilde{G}({\bf k},\mu_{0})$, which is the Laplace
transform of $G({\bf k},N_{0})$ with respect to $N_{0}$,  
can be written exactly
as \cite{Ma}:
\begin{eqnarray}
\tilde{G}({\bf k},\mu_{0})=\left(\mu_{0} +\frac{l^{2}}{6}k^{2}+\Sigma({\bf k})\right)^{-1}\label{ft4}
\end{eqnarray}
where $\Sigma({\bf k})$ denotes the proper self energy. We now define a
trial-Hamiltonian ${\cal H}$:
\begin{eqnarray}
\beta {\cal H}[\vec{\psi}]=\frac{1}{2}\int_{{\bf k}}\vec{\psi}(-{\bf k})\tilde{{\cal
    G}}^{-1}({\bf
  k},\mu_{0})\vec{\psi}({\bf k})\label{ft6}
\end{eqnarray}
where $\tilde{{\cal G}}({\bf k},\mu_{0})$ is an approximate correlation
function with an approximate proper self-energy ${\cal M}({\bf k})$:
\begin{eqnarray}
\tilde{{\cal G}}({\bf k},\mu_{0})=\left(\mu_{0} +\frac{l^{2}}{6}k^{2}+{\cal M}({\bf k})\right)^{-1}.\label{ft4a}
\end{eqnarray}
In this notation the well-known Feynman inequality is given by:
\begin{eqnarray}
F\le {\cal F}+\langle H-{\cal H}\rangle_{{\cal H}}\label{ft8}
\end{eqnarray}
where
\begin{eqnarray}
\langle\dots\rangle_{{\cal H}}=\lim_{n \to 0}\frac{\int{\cal
    D}\vec{\psi}\,\dots\exp\{-\beta {\cal H}\}}{\int{\cal
    D}\vec{\psi}\,\exp\{-\beta {\cal H}\}}\label{ft9}
\end{eqnarray}
is the mean-value and ${\cal F}$ the free energy with respect to ${\cal H}$.
 The right hand side of the inequality
(\ref{ft8}) has to be minimized with respect to 
${\cal M}$, which is our variational function.
The general minimization condition reads
\begin{eqnarray}
\frac{\delta}{\delta {\cal M}({\bf
    q})}({\cal F}+\langle H-{\cal H}\rangle_{{\cal H}})=0\label{ft13}
\end{eqnarray}
where $\delta/\delta{\cal M}({\bf q})$ denotes the functional derivative with
respect to ${\cal M}({\bf q})$. This condition (\ref{ft13}) is equivalent to \cite{HaWiVi}
\begin{eqnarray}
{\cal M}({\bf q})&=&\frac{2\pi bz^{2} n}{\kappa^{2}}\int_{\bf k}\frac{1}{\mu_{0}
  +\frac{l^{2}}{6}k^{2}+{\cal M}({\bf k})}\nonumber \\
&+&4\pi bz^{2}\int_{{\bf k}}\frac{1}{\left(\kappa^{2}+({\bf q}+{\bf k})^{2}\right)\left( \mu_{0}
  +\frac{l^{2}}{6}k^{2}+{\cal M}({\bf k})\right)}\label{ft14}
\end{eqnarray}
This is a non-linear integral equation for ${\cal M}({\bf q})$, which in the
following has to
be solved approximately, since the exact solution is unknown. At this point it
should be stressed, that Eq. (\ref{ft14}) represents the well known Hartree
approximation. As our aim is to get a force-size relationship according to
Eqs. (\ref{model4}) and (\ref{model5}) the ansatz for the approximate proper
self energy
\begin{eqnarray}
{\cal M}({\bf q})={\cal M}({\bf 0})+\alpha {\bf k}^{2}+{\cal O}({\bf k^{4}})\label{neu1}
\end{eqnarray}
is valid for small external forces $f$. The constant ${\cal M}({\bf 0})$ is
absorbed by the shifted chemical potential $\mu=\mu_{0}+{\cal M}({\bf 0})$.
Consequently the only task is to calculate the coefficient $\alpha$
self-consistently from Eq. (\ref{ft14}). This is possible for small inverse
screening length $\kappa$, which means in the long ranged limit of the
Debye-H\"uckel potential. The result is:
\begin{equation}
\alpha=\frac{2bz^{2}}{3\pi \mu \kappa}
\left [1+{\cal O}\left (\kappa l \over \sqrt{\mu}  \right)\right]\label{s6}
\end{equation}
For details see \cite{HaWiVi}.

\subsection{Results}
Let us describe the consequences  briefly.
Inserting the approximate result for the proper self-energy $\Sigma_{}({\bf k})$ into
Eq. (\ref{ft4}) yields an explicit expression for the correlation function
$\tilde{G}({\bf k},\mu)$ with the shifted chemical potential $\mu$ as
mentioned above
\begin{equation}
\tilde{G}({\bf k},\mu)=\left( \mu
  +\frac{l^{2}}{6}k^{2}+\frac{2bz^{2}}{3\pi  \mu \kappa}k^{2}\right)^{-1}\label{r1}
\end{equation}
Now the conformational free energy of the chain under the influence of a force
${\bf f}$ can be calculated very easily by
$\phi(\mu,{\bf f})=-\ln {\tilde G}({\bf 0},\mu;{\bf f})=-\ln {\tilde G}(-i\beta{\bf f},\mu)$. Using the
well-known thermodynamic relationship
\begin{eqnarray}
N=\frac{\partial \phi(\mu,{\bf f})}{\partial \mu}=-\frac{\partial \ln {\tilde G}(-i\beta{\bf f},\mu)}{\partial \mu}\label{r2}
\end{eqnarray}
we express $\mu$ depending on its conjugate variable $N$ and the force ${\bf
  f}$. Note that the variable $N$ is not the bare number of monomers $N_{0}$ since
$\mu$ is a shifted chemical potential, but $N$ is proportional to
$N_{0}$, indeed it is easily seen
that $N < N_0$. Considering only singular terms in $\kappa$ and neglecting terms of 
order $f^{4}$ this calculation yields:
\begin{eqnarray}
\mu=\frac{1}{N}+\frac{4Nbz^{2}\beta^{2}f^{2}}{3\pi \kappa}+{\cal O}(f^{4})\label{r2a}
\end{eqnarray}
Substituting Eq. (\ref{r2a}) into Eq. (\ref{r1}) we get $G({\bf k},N)$.
According to Eqs. (\ref{model4}) and (\ref{model5}) $\langle{\bf R}^2_\parallel\rangle$ and
$\langle{\bf R}^2_\perp\rangle$ can be calculated from Eq. (\ref{r2}). Expanding in a
power series for small forces to second order and again considering only
most singular terms  for small $\kappa$, 
$\langle{\bf R}^2_\parallel\rangle$ becomes for $\beta f/\kappa<1$
\begin{eqnarray}
\langle{\bf R}^2_\parallel\rangle[f]=\frac{4N^{2}bz^{2}}{3\pi
  \kappa}+\beta^{2} f^{2}\frac{8N^{4}b^{2}z^{4}}{9 \pi^{2} \kappa^{2}}+{\cal O}(\frac{\beta^{4}f^{4}}{\kappa^{4}})\label{r4}
\end{eqnarray}
 The square root of the mean square elongation  can
be written according to Eq. (\ref{r4}) as
\begin{equation}
\sqrt{\langle{\bf R}^2_\parallel\rangle[f]-\langle{\bf R}^2_\parallel\rangle[0]}\sim f\label{r6}
\end{equation}
which is a Hook-like law.

Using the same approximation as 
mentioned above the root mean square end-to-end
distance 
perpendicular  to the force $f$, $\sqrt{\langle{\bf R}^2_\perp\rangle}$, decreases with
$f$ for $\beta f/\kappa<1$, which is contrary to a Gaussian chain \cite{degennes}.
In
particular: 
\begin{equation}
\langle{\bf R}^2_\perp\rangle[f]=\frac{8N^{2}bz^{2}}{3\pi
  \kappa}-\beta^{2} f^{2}\frac{16N^{4}b^{2}z^{4}}{3 \pi^{2} \kappa^{2}}+{\cal O}(\frac{\beta^{4}f^{4}}{\kappa^{4}})\label{r5}
\label{r7}
\end{equation}
For  $f=0$
the 
perpendicular end-to-end distance squared becomes doubles exactly $\langle{\bf
  R}^2_\parallel\rangle[0]$, since we have assumed rotational isotropy. 

The latter relation ship for zero forces does not agree with one of the two
scaling limits, i.e., strong polyelectrolyte regime or excluded volume regime
(large $\kappa$). In the strong polyelectrolyte regime, very small $\kappa$ the
chain size does not depend on the screening parameter \cite{degennesblob},
whereas here it does. The reason is that we are here in the intermediate
regime where $1/\kappa$ is of the same order as the chain size. This is
visible by a detailed study of the renormalization in the chain
length from $N_{0}$ to $N$. It is exactly this regime which we need to study
the statistical mechanics of networks.

\section{Polyelectrolyte Networks}
In this section we calculate the modulus of a polyelectrolyte network with a
random crosslink configuration. 

\subsection{Model}

A network is formed out of a very long chain by the instantaneous introduction
of a sufficiently large number of
crosslinks in the liquid phase (see for example \cite{PanRab}).
We restrict ourselves to a network, which consists of flexible, weakly charged 
strands. Consequently the Edwards model is an appropriate tool to describe the
network \cite{Ledwards}. Therefore the underlying Hamiltonian is the same as
for the single chain Eq. (\ref{model1}). Again we assume Debye-H\"uckel
interaction between the monomers.
 For the introduction of
crosslinks
we choose the standard way suggested by Deam and Edwards \cite{deam}. We
assume for mathematical convenience four functional crosslinks which join
two arbitrary segments ${\bf r}(s_{i})$ and ${\bf r}(s_{j})$ along the
chain.
Of course, the value for the free energy then depends on the specific choice of
the pairs of monomers, but on macroscopic scale only the statistical average
on any crosslink configuration
is of importance. Nevertheless this requires non - Gibbsian statistical
mechanics in the sense that the crosslink positions represent quenched degrees
of freedom.

To calculate therefore the free energy $F$ of the network, we have to take the
statistical average over all crosslink configurations ${\bf S}$. This
represents the fact, that $F$ is a self averaging quantity.
\begin{eqnarray}
F(N_{\rm tot},N_{\rm c})=-k_{\rm B}T\int {\mbox d}{\bf S}\, {\cal P}({\bf S})\ln Z({\bf
  S}).\label{m2}
\end{eqnarray}
$Z({\bf S})$ is the constrained partition function for a network with
the crosslink configuration ${\bf S}$ (see section II), $N_{\rm c}$ is the number of crosslinks and ${\cal P}({\bf S})$ is the crosslink
distribution function. Since we assume that the crosslinks are instantaneously
introduced in the liquid phase, ${\cal P}({\bf S})$ is yielded by the
constrained partition function of the liquid phase $Z^{(0)}({\bf S})$ Eq. (\ref{frh1}).
 The crosslink distribution function ${\cal P}({\bf S})$
is simply:
\begin{eqnarray}
{\cal P}({\bf S})=\frac{Z^{(0)}({\bf S})}{\int {\mbox d}{\bf S}^{'} Z^{(0)}({\bf
  S}^{'})}\,.\label{m4}
\end{eqnarray}
In the following it appears to be reasonable to 
assume that the so chosen distribution function does not depend on the specific
deformation of the network. 
Note that $Z(\bf S)$ differs generally
from $Z^{(0)}({\bf S})$. Since we are interested
in deformations of the network,  $Z(\bf S)$ is the partition function of the
deformed network.

To calculate the free energy $F$ (Eq. (\ref{m2})) explicitly it is convenient
to  use of the so called replica
trick \cite{deam}. Define
\begin{eqnarray}
F_{m}(N_{\rm tot},N_{\rm c})=-k_{\rm B}T\ln \int {\mbox d}{\bf S}\,Z^{(0)}({\bf
S})Z^{m}({\bf S})\,,\label{m5}
\end{eqnarray}
where $m$ is the Replica index. The free energy $F$, which is averaged  over the disorder of the crosslinks, reads \cite{deam}:
\begin{eqnarray}
F(N_{\rm tot},N_{\rm c})=\frac{{\mbox d}F_{m}(N_{\rm tot},N_{\rm c})}{{\mbox
    d}m} {\Bigg\vert}_{m=0} \label{ef6}
\end{eqnarray}
As in the case of the single chain, the free energy $F$ is calculated by making use of
its relation to the corresponding distribution functions and Green functions
of the corresponding propagator (see \cite{HaWiVi} for the technical details).

\subsection{Field-theoretical Formulation and Results}

As in the case of the single chain the external force acts on the ends of the
chain and is reintroduced by the analytic continuation to the complex plane:
\begin{eqnarray}
\tilde{G}({\bf \hat{k}},\mu_{0},z_{\rm c},{\bf
  f})=\tilde{G}({\bf k}^{(0)},{\bf k}^{(1)}-i\beta {\bf f},\dots,{\bf k}^{(m)}-i\beta
{\bf f},\mu,z_{\rm c},{\bf f}={\bf 0})\label{ft1a}
\end{eqnarray} 
Here $\tilde{G}({\bf \hat{k}},\mu_{0},z_{\rm c},{\bf f})$ is the Greens-function,
 where ${\bf
  \hat{k}}$ is the wave vector in the $3(m+1)$-dimensional Fourier transformed
  replica space and ${\bf k}^{(i)}$ is the three-dimensional wave vector of the
  $i$-th replica, $\mu_{0}$
is the chemical potential of the monomers and $z_{\rm c}$ is the fugacity of the
crosslinks.
To calculate the Greens function $\tilde{G}({\bf \hat{k}},\mu_{0},z_{\rm c})$ we
introduce a field theory. The field theoretical forceless Hamiltonian for the
network in Fourier space reads \cite{PanRab}, \cite{HaWiVi}:
\begin{eqnarray}
H[\vec{\psi}({\bf \hat{q}})]&=&\int_{{\bf
    \hat{q}}}\left[\frac{\mu}{2}\vec{\psi}({\bf \hat{q}})\vec{\psi}(-{\bf
    \hat{q}})+\frac{l^{2}}{2}{\bf \hat{q}}^{2}\vec{\psi}({\bf
    \hat{q}})\vec{\psi}(-{\bf \hat{q}})\right]\nonumber \\
&-&\frac{z_{\rm c}}{8}\int_{{\bf \hat{q}}_{1},{\bf \hat{q}}_{2},{\bf
    \hat{q}}_{3},{\bf \hat{q}}_{4}} \vec{\psi}({\bf
  \hat{q}}_{1})\vec{\psi}({\bf \hat{q}}_{2})\vec{\psi}({\bf
  \hat{q}}_{3})\vec{\psi}({\bf \hat{q}}_{4})\delta({\bf \hat{q}}_{1}+{\bf
  \hat{q}}_{2}+{\bf \hat{q}}_{3}+{\bf \hat{q}}_{4})\nonumber \\
&+&\sum_{k=0}^{m}\left[\int_{{\bf \hat{q}}_{1},{\bf
      \hat{q}}_{2}}\vec{\psi}({\bf \hat{q}}_{1})\vec{\psi}({\bf \hat{q}}_{2})\prod_{l\neq
    k}\delta({\bf q}_{1}^{(l)}+{\bf q}_{2}^{(l)})\right]\label{ft2} \\
&\times&\left[\int_{{\bf \hat{q}}_{3},{\bf \hat{q}}_{4}}\vec{\psi}({\bf
    \hat{q}}_{3})\vec{\psi}({\bf \hat{q}}_{4})\prod_{l\neq
    k}\delta({\bf q}_{3}^{(l)}+{\bf q}_{4}^{(l)})\right]V^{(k)}({\bf
  q}_{3}^{(k)}+{\bf q}_{4}^{(k)})\delta({\bf q}_{1}^{(k)}+{\bf
  q}_{2}^{(k)}+{\bf q}_{3}^{(k)}+{\bf q}_{4}^{(k)})\nonumber
\end{eqnarray}
where $V^{(k)}(q)$ is the Fourier transform of the Debye-H\"uckel potential in
the $k$-th replica and $\vec{\psi}$ an $n$-component vector field. The further procedure of the calculation is exactly
analogous to the calculation concerning the single chain presented
above. Again we introduce a Gaussian trial Hamiltonian with the correlation
function:
\begin{eqnarray}
\tilde{{\cal G}}({\bf \hat{k}},\mu_{0},z_{\rm c})=\left(\mu_{0}
  +\frac{l^{2}}{6}\hat{k}^{2}+{\cal M}({\bf \hat{k}},z_{\rm c})\right)^{-1}\label{ft4}
\end{eqnarray}
Then we make use of
Feynman's inequality for a variational principle. The variational parameter
again is the approximate proper self energy ${\cal M}({\bf \hat{k}},z_{\rm c})$. The only difference is, that now
the proper self energy depends on a {\bf replicated} Fourier vector ${\bf {\hat
  k}}$. For details see \cite{wilder}. Assuming a replica symmetry this procedure leads to the following
self-consistent equation for $M({\bf \hat{k}},z_{\rm c})$ 
\begin{eqnarray}
M({\bf \hat{k}},z_{\rm c})&=&\frac{2\pi
  bz^{2}n(m+1)V_{0}^{m}}{\kappa^{2}}\int_{{\bf
    \hat{q}}}\,\frac{1}{\mu+\frac{l^{2}}{6}{\bf \hat{q}}^{2}+M({\bf
    \hat{q}},z_{\rm c})}\nonumber \\
&+&4\pi bz^{2}\int_{{\bf q}^{(0)}}\,\frac{1}{({\bf q}^{(0)}+{\bf
    k}^{(0)})^{2}+\kappa^{2}}\frac{1}{\mu+\frac{l^{2}}{6}({\bf q}^{(0)},{\bf
    k}^{(1\dots m)})^{2}+M({\bf q}^{(0)},{\bf k}^{(1\dots m)})}\label{ft12}\\
&+&4\pi bz^{2}m\int_{{\bf q}^{(1)}}\,\frac{1}{({\bf q}^{(1)}+{\bf
    k}^{(1)})^{2}+\kappa^{2}}\frac{1}{\mu+\frac{l^{2}}{6}({\bf k}^{(0)},{\bf
    q}^{(1)},{\bf k}^{(2\dots m)})^{2}+M({\bf k}^{(0)},{\bf q}^{(1)},{\bf k}^{(2\dots m)})}\nonumber \\
&+&
\frac{z_{\rm c}}{2}(n+2)\int_{{\bf \hat{q}}}\,\frac{1}{\mu+\frac{l^{2}}{6}{\bf
    \hat{q}}^{2}+M({\bf \hat{q}},z_{\rm c})},\nonumber
\end{eqnarray}
where ${\bf k}^{(i\dots m)}=({\bf k}^{(i)},\dots ,{\bf k}^{(m)})$.
This equation has to be solved approximately. 
In analogy to the calculations on the single chain \cite{HaWiVi} we restrict
ourselves to small external forces applied on the ends of the crosslinked
chain. Therefore we make the same ansatz for the proper-self energy as in the
case of the single chain. The only difference is that the proper
self energy in this paper is a function depending on the replica-space wave
vector ${\bf \hat{q}}$:
\begin{equation}
M({\bf \hat{q}})=a_{0}+a_{1}{\bf \hat{q}}^{2}+{\cal O}({\bf \hat{q}}^{4})\label{l1}
\end{equation}
In the limit of a weakly charged network and in the long ranged limit of the
Debye-H\"uckel potential the result of the self-consistency calculation
according to Eq. (\ref{ft12}) for $a_{0}$ and $a_{1}$ is (for details see \cite{wilder}):
\begin{equation}
a_{1}=\frac{2bz^{2}}{3l^{2}\pi \mu_{\rm r} \kappa}
\left [1+{\cal O}\left (\kappa l \over \sqrt{\mu_{\rm r}}  \right)\right]\label{l3}
\end{equation}
and
\begin{eqnarray}
a_{0}&=&\frac{\kappa}{\mu}\left(\frac{2bz^{2}(m+1)}{\pi}-\frac{bz^{2}(m+1)}{2}+\frac{2\kappa^{2}l^{2}bz^{2}(m+1)}{9\mu
    \pi}-\frac{\kappa^{2}l^{2}bz^{2}(m+1)}{12\mu}\right)\\ \nonumber
&-&\frac{\kappa}{\mu}\left(\frac{2\kappa^{2}z_{\rm
      c}m}{3\pi^{2}}-\frac{\kappa^{2}z_{\rm c}\gamma
    m}{4\pi^{2}}-\frac{\kappa^{2}z_{\rm c}\ln(\kappa
    l/\sqrt{\pi})m}{2\pi^{2}}-\frac{\kappa^{2}z_{\rm c}}{6\pi^{2}}\right)\label{l9}
\end{eqnarray}
Therefore the approximate Greens function reads:
\begin{equation}
\tilde{\cal G}({\bf \hat{k}},\mu,z_{\rm c})=\frac{1}{\mu+a_{0}+\frac{l^{2}}{6}{\bf
    \hat{k}}^{2}+a_{1}{\bf \hat{k}}^{2}}\label{l11}
\end{equation}
From Eq. (\ref{l11}) the grand canonical partition function in replica space
under the influence of an external constant force $f$ acting on the ends of
the chain can be calculated:
\begin{eqnarray}
\Xi_{m}(\mu,z_{\rm c},{\bf f})&=&\tilde{\cal G}({\bf k}^{(0)},{\bf k}^{(1)}-i\beta {\bf
  f},\dots,{\bf k}^{(m)}-i\beta
{\bf f},\mu,z_{\rm c})\arrowvert_{{\bf \hat{k}}={\bf 0}}\nonumber \\
&=&\frac{1}{\mu+a_{0}-\frac{l^{2}}{6}m\beta^{2}{\bf f}^{2}-a_{1}m\beta^{2}{\bf
    f}^{2}}\label{erg1}
\end{eqnarray}
Here we reintroduced the force $f$ according to Eq. (\ref{ft1a}). The monomer
chemical potential and the fugacity of cross-links which parameterize the grand
canonical partition function in Eq. (\ref{erg1}), should be expressed in terms
of the parameters $N_{\rm tot}$ and $N_{\rm c}$. According to Panyukov and Rabin
\cite{PanRab} the expression $F_{m}(N_{\rm tot},N_{\rm c})$ can be calculated by the
method of steepest descent in the thermodynamic limit $N_{\rm tot}, N_{\rm c}\to
\infty$:
\begin{eqnarray}
F_{m}(N_{\rm tot},N_{\rm c})/k_{\rm B}T=-\ln\Xi(\mu,z_{\rm c})-N_{\rm
  tot}\mu+N_{\rm c}\ln z_{\rm c}\label{erg2}
\end{eqnarray}
Consequently the fugacity $z_{\rm c}$ of cross-links and the chemical potential
$\mu$ of monomers can be obtained by minimizing the right-hand side of Eq. (\ref{erg2}): 
\begin{eqnarray}
N_{\rm tot}=-\frac{\partial\ln\Xi_{m=0}(\mu,z_{\rm c})}{\partial\mu}\label{erg4}
\end{eqnarray}
and
\begin{eqnarray}
N_{\rm c}=\frac{\partial\ln\Xi_{m=0}(\mu,z_{\rm c})}{\partial\ln z_{\rm c}}\label{erg5}
\end{eqnarray}
in the limit of a vanishing replica index $m$. In the weakly charged limit of
the network and in the long-ranged limit of the Debye-H\"uckel potential the
result for $\mu$ and $z_{\rm c}$ is:
\begin{eqnarray}
\mu={\bar N}\label{neu5}
\end{eqnarray}
and
\begin{eqnarray}
z_{\rm c}=\frac{6\pi^{2}\bar{N}^{2}}{\kappa^{3}}+\frac{3\pi(4-\pi)bz^{2}}{\kappa^{2}}\label{erg13}
\end{eqnarray}
 Moreover it can be shown
\cite{PanRab}, that the conformational free energy of the network is given by:
\begin{equation}
F(N_{\rm tot},N_{\rm c},f)=-k_{\rm B}T\frac{\partial\ln\Xi_{m}(\mu,z_{\rm c})}{\partial m}{\Bigg\vert}_{m=0}\label{erg40}
\end{equation}
If the free energy $F$ is known the force-size relationship is the simply 
calculated by
the derivative of $F$ with respect to the external force $f$:
\begin{equation}
\langle R \rangle
=-\frac{\partial F(N_{\rm tot},N_{\rm c},f)}{\partial f}\label{erg50}
\end{equation}
After having inserted the results for $z_{\rm c}$ and $\mu$ in the force size
relationship, expanded for small charges $z$ neglecting terms of order 
$z^{4}$ and higher, since only weakly charged networks are stable, and
considered only terms of leading order with respect to small $\kappa$ the
force-size relationship reads:
\begin{equation}
\langle R \rangle=\left(\frac{bz^{2}}{3\bar{N}^{2}\pi\kappa}+\frac{l^{2}}{6\bar{N}}\right)\beta f+{\cal O}(z^{4},\kappa)\label{erg14}
\end{equation}
This is again the result for the small deformation regime. The result
describes a Hookian law for the force extension relation and defines the
elastic modulus of the polyelectrolyte network.
Note further
that this result is valid for small forces, i.e. $\beta f/\kappa<1$. 
Therefore the modulus for the small screening and the low deformation
regime of the network reads:
\begin{equation}
G=\left(\frac{\beta bz^{2}}{3\bar{N}^{2}\pi\kappa}+\frac{\beta l^{2}}{6\bar{N}}\right)^{-1}\label{erg15}
\end{equation}
 The modulus depends on the density
of the crosslink and on the Debye screening parameter. Thus both contributions
enter in a significant way. 
Most important  is  that part of the modulus stemming from the interactions  
depends on the crosslink density $\bar{N}$ squared. This now indicates that in
the case of strongly interacting soft materials the FRH is no longer
applicable. The modulus computed here is a strong combination of both
contributions. Moreover we stress that the result is beyond perturbation
theory and RPA - type approximations in contrast to earlier suggestions
published previously \cite{muthu}. The author in this paper suggests a modulus
which is not changed by the interactions.

\section{Swelling Behavior}

Let us finally discuss the consequences  of the above results to the swelling
behavior of a simple polyelectrolyte network. The general theory of swelling
for these systems is not simple and will be studied in a subsequent
publication. Nevertheless we have a good basis to study the general swelling
behavior of polyelectrolyte networks. To do so we assume that the equilibrium
swelling is given by the $c^{*}$ - network \cite{degennes}. The consequence of
this assumption is that the single chain behavior can be used for the
determination of the degree of swelling. Thus we may restrict ourselves on one
mesh of the network, and we can use the single chain behavior studied in the
first section of the paper. Of course, we have to neglect complications by the
presence of entanglements and topological restrictions. Therefore we can apply
the following results only for mesh sizes below typical entanglement
lengths. The strong electrostatic repulsion, however, does  not allow for a
large degree of topological restrictions, which makes the assumption of no
entanglements reasonable. To estimate the equilibrium degree of swelling, we
use an ideal $c^{*}$ - network whose monodisperse meshes consist of chains
with  $N$ monomers. These chains are connected by the process of
endlinking, where the functionality of an end of the chain should be six in
three dimensions. Thus in the ideal totally stretched state of the network the
end-links are on the lattice points of a simple cubic lattice with the lattice
constant $Nl$. Figure 1 shows such a swollen network in two dimensions.
\begin{figure}
\epsfig{file=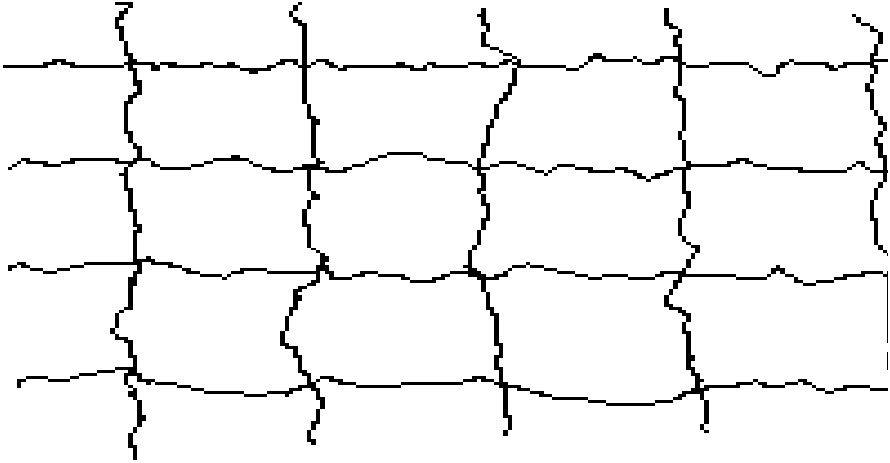,width=14cm,height=8cm}
\caption{Swollen network in two dimensions}
\end{figure}
For long chains this ideal network is of course not very realistic. On the one
hand entanglements occur, which disturb the ideal structure drastically. On the
other hand two points of the network could be connected, which would be in the
ideal, swollen state far away from each other. If the network, however, is
synthesized out of short chains, that means $N$ is small, then these defects
should occur with a much smaller probability. Consequently in this section we
have to restrict ourselves to networks, which are made of very short chains.

In the dry state without having contact to a solvent the
considered network is a dense system of neutral polymers, because counterions
are bound to the charges on the polymers. In solvent some counterions
dissociate and as a result we get a screened Coulomb interaction between the
charges, which swells the network. To describe the swelling of a network
correctly, one has to know the complete free energy of all the states 
between unswollen and
swollen. This means that we have to have a theory that is valid for the
network from the unswollen to the swollen state. As far as we know such a
description does not exist by now, because theoretical considerations are only
 made in special physical limits. Therefore in this section we use the idea of
 the $c^{*}$ - network: we compare the volume of the dry and
the swollen state of the above mentioned model-network.

The dry state is a dense
system of neutral polymers. which means that the remaining excluded volume
interaction is screened \cite{doi}. Consequently the average end-to-end
distance of such a chain is:
\begin{eqnarray}
\langle R_{0}^{2} \rangle \propto l^{2}N\label{neu2}
\end{eqnarray} 
We assume that the considered model network is formed in an equilibrium state,
where the mean end-to-end distances of the chains obey Eq. (\ref{neu2}). In
the swollen state the averaged end-to-end distances of the chains connecting
two neighboring endlink-points in the network can be calculated from
Eqs. (\ref{r4}) and (\ref{r5}). For vanishing external forces $f$:
\begin{eqnarray}
\langle R^{2} \rangle \propto \frac{N^{2}bz^{2}}{\kappa}\label{neu3}
\end{eqnarray}
Thus for our model network we can introduce a parameter of swelling $Q$
defined as $Q=V/V_{0}$, where $V$ is the volume of the swollen network and
$V_{0}$ is the volume of the unswollen network:
\begin{eqnarray}
Q=\frac{V}{V_{0}}=
\left(
\frac{\langle R^{2} \rangle}{\langle R_{0}^{2} \rangle}\right)^{\frac{3}{2}}
\propto \frac{b^{3/2}z^{3}}{\kappa^{3/2}l^{3}{\bar N}^{3/2}}\label{neu4}
\end{eqnarray}
Here ${\bar N}=N_{\rm e}/N_{\rm tot}$ 
is the number of endlinks $N_{\rm e}$ divided by the
total number of monomers $N_{\rm tot}$, i. e. the endlink-density. In our case
$N=1/{\bar N}$ is the averaged number of monomers per strand.

Here we see clearly that in contrast to neutral networks the degree of
swelling $Q$ does not only depend on the crosslink density, but also on the
range of the interactions via $\kappa$ and therefore the ionization. A typical
dependence on the different quantities is shown in figure 2.
\begin{figure}
\epsfig{file=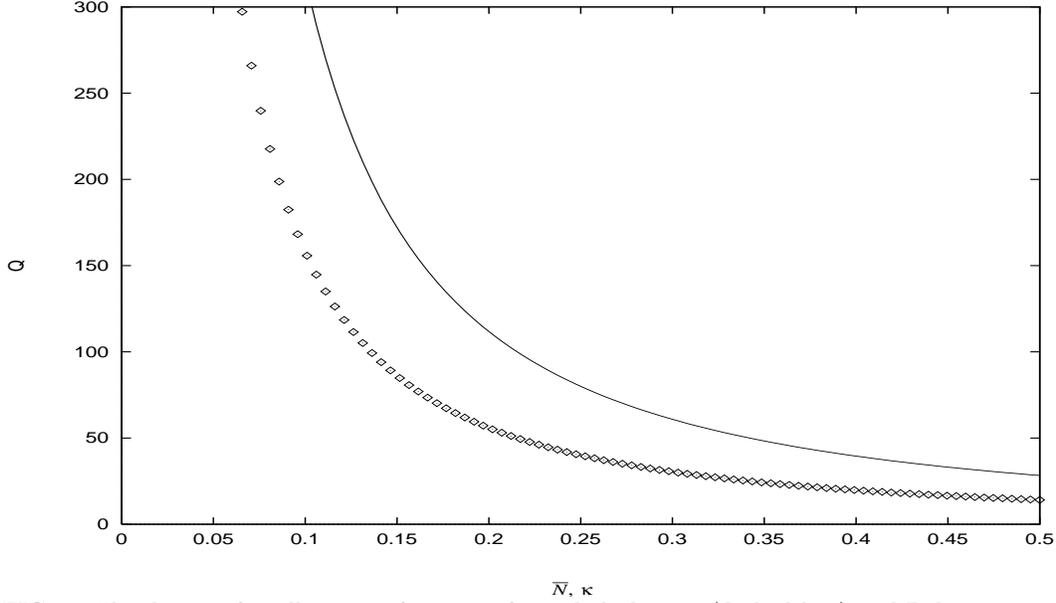,width=14cm,height=8cm}
\caption{The degree of swelling as a function of crosslink density (dashed line) and Debye
  screening parameter (bold-faced line).}
\end{figure}

\section{Discussion}
In this paper we developed different stages of the problem of the elastic
behavior of polyelectrolyte networks. This problem required a new approach and
the classical theories for neutral networks cannot be used, so far, 
except in the
limit of large salt concentration, i.e., $\kappa \to \infty$. 
In this regime the
interaction potential between the segments is very short ranged and the
network behaves roughly as an "effective excluded volume network". Naturally
this regime is not very interesting, since no real new insight can be expected
\cite{muthu}. On the other hand, the small $\kappa$ regime is very important
and novel theories can be formulated \cite{tanniestapp}. In this regime we
found the strong invalidity  of the Flory - Rehner - hypothesis, which states
that different parts of the free energy can be added. The elastic modulus of
the network contains both: the entropic part and the interactions. Both seem to
show up as two springs in a series. The equilibrium swelling is determined by
interactions and crosslink density.

\section*{Acknowledgments}

The authors wish to thank 
 Firma Stockhausen Gmbh, D-47705 Krefeld, Germany for financial support.

\begin{appendix}
\section{}
Starting from Eq. (\ref{frh2}) we introduce the densities $\rho_{\rm p}({\bf
  k})$ of the polymers and $\rho_{\rm i}({\bf k})$ of the counterions. The
result is:
\begin{eqnarray}
Z(\rho_{\rm p},\rho_{\rm i},{\bf S})&=&\frac{1}{N_{\rm i}!}\int {\cal D}{\bf r}(s)\int
\frac{{\mbox d}{\bf R}_{1}}{V_{0}}\dots\int \frac{{\mbox d}{\bf R}_{N_{\rm
      i}}}{V_{0}} \prod_{\bf k} \delta\left(\rho_{\rm p}({\bf k})-\frac{1}{V_{0}}\int {\mbox d}s\,{\rm
  e}^{i{\bf kr}(s)}\right)\delta\left(\rho_{\rm i}({\bf k})-\frac{1}{V_{0}}\sum_{j=1}^{N_{\rm
    i}}{\rm e}^{i{\bf kR}_{j}}\right)\nonumber \\
&\times&\exp\left[-{3 \over 2l^{2} } \int^{N_{0}}_{0} \mbox{d} s \
\left({\mbox{d} {\bf r} \over \mbox{d}
    s}\right)^2-\ln\left(\prod_{(i,j)}\delta[{\bf r}(s_{i})-{\bf
    r}(s_{j})\right)-\frac{\beta}{2}\sum_{i,l=1}^{N_{i}}\frac{bz_{i}z_{l}}{{\bf
    k}^{2}}{\rm e}^{i{\bf k}({\bf R}_{i} - {\bf R}_{l})}\right]\nonumber\\
&\times&\exp\left[-\frac{\beta}{2}\int^{N_{0}}_{0}\mbox{d} s \int^{N_{0}}_{0}\mbox{d}
s'\frac{bz^{2}}{{\bf k}^{2}}{\rm e}^{i{\bf k}({\bf r}(s)-{\bf
    r}(s'))}-\beta\sum_{i=1}^{N_{i}}\int^{N_{0}}_{0}\mbox{d}
s\frac{bzz_{i}}{{\bf k}^{2}}{\rm e}^{i{\bf k}({\bf r} (s) - {\bf
    R}_{i})}\right]\label{afrh3}
\end{eqnarray}
with $b=e^{2}/4\pi
\epsilon_{0}\epsilon_{\rm r}k_{\rm B}T$  the Bjerrum-length. According to the $\delta$-distributions in Eq. (\ref{afrh3}) the partition
function can be rewritten as:
\begin{eqnarray}
Z(\rho_{\rm p},\rho_{\rm i},{\bf S})&=&\frac{1}{N_{\rm i}!}\int {\cal D}{\bf r}(s)\int
\frac{{\mbox d}{\bf R}_{1}}{V_{0}}\dots\int \frac{{\mbox d}{\bf R}_{N_{\rm
      i}}}{V_{0}} \prod_{\bf k} \delta\left(\rho_{\rm p}({\bf k})-\frac{1}{V_{0}}\int_{0}^{N_{0}} {\mbox d}s\,{\rm
  e}^{i{\bf kr}(s)}\right)\label{afrh4} \\
&\times&\delta\left(\rho_{\rm i}({\bf k})-\frac{1}{V_{0}}\sum_{j=1}^{N_{\rm
    i}}{\rm e}^{i{\bf kR}_{j}}\right)\exp\left[-{3 \over 2l^{2} } \int^{N_{0}}_{0} \mbox{d} s \
\left({\mbox{d} {\bf r} \over \mbox{d}
    s}\right)^2-\ln\left(\prod_{(i,j)}\delta[{\bf r}(s_{i})-{\bf
    r}(s_{j})\right)\right]\nonumber\\
&\times&\exp\left[-\frac{1}{2}\sum_{\bf k} V_{\rm pp}({\bf k})\rho_{\rm p}({\bf k})\rho_{\rm p}(-{\bf
    k})-\frac{1}{2}\sum_{\bf k} V_{\rm ii}({\bf k})\rho_{\rm i}({\bf
    k})\rho_{\rm i}(-{\bf k})-\sum_{\bf k} V_{\rm pi}({\bf
    k})\rho_{\rm p}({\bf k})\rho_{\rm i}(-{\bf k})\right]\nonumber
\end{eqnarray}
where $V_{\rm pp}({\bf k})=bz^{2}/{\bf k}^{2}$. $V_{\rm pi}$ and $V_{\rm ii}$
are defined analogously. The $\delta$-functions in Eq. (\ref{afrh4}) can be
rewritten by use of a $\phi$- and $\phi '$-field:
\begin{eqnarray}
Z(\rho_{\rm p},\rho_{\rm i},{\bf S})&=&\frac{1}{N_{\rm i}!}\int {\cal D}{\bf r}(s)\int
\frac{{\mbox d}{\bf R}_{1}}{V_{0}}\dots\int \frac{{\mbox d}{\bf R}_{N_{\rm
      i}}}{V_{0}}\int {\cal D}\phi({\bf k})\exp\left[i\phi({\bf k})\left(\rho_{\rm p}({\bf k})-\frac{1}{V_{0}}\int_{0}^{N_{0}}
    {\mbox d}s\,{\rm e}^{i{\bf kr}(s)}\right)\right]\nonumber \\
&\times&\int {\cal D}\phi'({\bf
  k})\exp\left[i\phi'({\bf k})\left(\rho_{\rm
      i}({\bf k})-\frac{1}{V_{0}}\sum_{j=1}^{N_{\rm i}}{\rm e}^{i{\bf
        kR}_{j}}\right)\right]\label{afrh5} \\
&\times&\exp\left[-{3 \over 2l^{2} } \int^{N_{0}}_{0} \mbox{d} s \
\left({\mbox{d} {\bf r} \over \mbox{d}
    s}\right)^2-\ln\left(\prod_{(i,j)}\delta[{\bf r}(s_{i})-{\bf
    r}(s_{j})\right)\right]\nonumber \\
&\times&\exp\left[-\frac{1}{2}\sum_{\bf k} V_{\rm pp}({\bf k})\rho_{\rm p}({\bf k})\rho_{\rm p}(-{\bf
    k})-\frac{1}{2}\sum_{\bf k} V_{\rm ii}({\bf k})\rho_{\rm i}({\bf
    k})\rho_{\rm i}(-{\bf k})-\sum_{\bf k} V_{\rm pi}({\bf
    k})\rho_{\rm p}({\bf k})\rho_{\rm i}(-{\bf k})\right]\nonumber
\end{eqnarray}
Within a Gaussian approximation the partition function reads:
\begin{eqnarray}
Z(\rho_{\rm p},\rho_{\rm i},{\bf S})&=&\frac{1}{N_{\rm i}!}\int {\cal D}{\bf r}(s)\int
\frac{{\mbox d}{\bf R}_{1}}{V_{0}}\dots\int \frac{{\mbox d}{\bf R}_{N_{\rm
      i}}}{V_{0}}\int {\cal D}\phi({\bf k}){\rm e}^{i\phi({\bf k})\rho_{\rm
  p}({\bf k})}\nonumber \\
&\times&\left(1-i\phi({\bf k})\int_{0}^{N_{0}} {\mbox d}s\,{\rm e}^{i{\bf k}{\bf
      r}(s)}-\frac{\phi^{2}({\bf k})}{2}\int_{0}^{N_{0}} {\mbox d}s \int_{0}^{N_{0}} {\mbox d}s'\,{\rm
    e}^{i{\bf k}({\bf r}(s)-{\bf r}(s'))}\right)\nonumber \\
&\times&\int {\cal D}\phi'({\bf k}){\rm e}^{i\phi '{\bf k}\rho_{i}({\bf
    k})}\left(1-i\phi '({\bf k})\sum_{j=1}^{N_{\rm i}} {\rm e}^{i{\bf k}{\bf
      R}_{j}}-\frac{\phi^{2}({\bf k})}{2}\sum_{j,l=1}^{N_{\rm i}}{\rm
    e}^{i{\bf k}({\bf R}_{j}-{\bf R}_{l})}\right)\nonumber \\ 
&\times&\exp\left[-{3 \over 2l^{2} } \int^{N_{0}}_{0} \mbox{d} s \
\left({\mbox{d} {\bf r} \over \mbox{d}
    s}\right)^2-\ln\left(\prod_{(i,j)}\delta[{\bf r}(s_{i})-{\bf
    r}(s_{j})\right)\right]\label{afrh6} \\
&\times&\exp\left[-\frac{1}{2}\sum_{\bf k} V_{\rm pp}({\bf k})\rho_{\rm p}({\bf k})\rho_{\rm p}(-{\bf
    k})-\frac{1}{2}\sum_{\bf k} V_{\rm ii}({\bf k})\rho_{\rm i}({\bf
    k})\rho_{\rm i}(-{\bf k})-\sum_{\bf k} V_{\rm pi}({\bf
    k})\rho_{\rm p}({\bf k})\rho_{\rm i}(-{\bf k})\right]\nonumber
\end{eqnarray}
Denote by ${\tilde S}_{0}({\bf k})$ and $C_{0}({\bf k})$:
\begin{eqnarray}
{\tilde S}_{0}({\bf k})=\left<\int_{0}^{N_{0}} {\mbox d}s \int_{0}^{N_{0}}{\mbox d}s'\,{\rm
    e}^{i{\bf k}({\bf r}(s)-{\bf r}(s'))}\right>_{0}\label{afrh7}
\end{eqnarray}
and
\begin{eqnarray}
C_{0}({\bf k})=\left<\sum_{j,l=1}^{N_{\rm i}}{\rm
    e}^{i{\bf k}({\bf R}_{j}-{\bf R}_{l})}\right>_{0}\label{afrh8}
\end{eqnarray}
where $<\dots>_{0}$ means the expectation value with respect to $H_{0}$ with:
\begin{eqnarray}
\beta H_{0}={3 \over 2l^{2} } \int^{N_{0}}_{0} \mbox{d} s \
\left({\mbox{d} {\bf r} \over \mbox{d}
    s}\right)^2+\ln\left(\prod_{(i,j)}\delta[{\bf r}(s_{i})-{\bf
    r}(s_{j})\right)\label{afrh9}
\end{eqnarray}
Line two and three of Eq. (\ref{afrh6}) are now replaced by the mean values of
the terms
concerning $H_{0}$. The result is considered as the first terms of a series
expansion of an exponential function. Consequently the exponential function is
reintroduced. After having integrated over $\phi$ and $\phi'$ the partition
function reads \cite{vibebe}:
\begin{eqnarray}
Z(\rho_{\rm p},\rho_{\rm i},{\bf S})&=&\exp\left[-\frac{1}{2}\sum_{\bf
    k}\left(\frac{1}{{\tilde S}_{0}({\bf k})}+V_{\rm pp}({\bf
      k})\right)\rho_{\rm p}({\bf k})\rho_{\rm p}(-{\bf k})\right]\nonumber \\
&\times&\exp\left[-\frac{1}{2}\sum_{\bf
    k}\left(\frac{1}{C_{0}({\bf k})}+V_{\rm ii}({\bf
      k})\right)\rho_{\rm i}({\bf k})\rho_{\rm i}(-{\bf k})-\sum_{\bf
    k}V_{\rm pi}({\bf k})\rho_{\rm p}({\bf k})\rho_{\rm i}(-{\bf
    k})\right]\label{afrh10} 
\end{eqnarray}
\end{appendix}

\end{document}